# The Higgs Factory Muon Collider Superconducting Magnets and Their Protection Against Beam Decay Radiation

N.V. Mokhov,[a] V.V. Kashikhin, S.I. Striganov, I.S. Tropin, and A.V. Zlobin

*Fermi National Accelerator Laboratory,*
*Batavia, IL 60510-5011, USA*
[a] *E-mail:* `mokhov@fnal.gov`

ABSTRACT: A low-energy medium-luminosity Muon Collider (MC) is being studied as a possible Higgs Factory (HF). Electrons from muon decays will deposit more than 300 kW in superconducting magnets of the HF collider ring. This imposes significant challenges to superconducting (SC) magnets used in the MC storage ring (SR) and interaction regions (IR). Magnet designs are proposed which provide high operating gradient and magnetic field in a large aperture to accommodate the large size of muon beams (due to low $\beta^*$), as well as a cooling system to intercept the large heat deposition from the showers induced by decay electrons. The distribution of heat deposition in the MC SR lattice elements requires large-aperture magnets in order to accommodate thick high-Z absorbers to protect the SC coils. Based on the developed MARS15 model and intensive simulations, a sophisticated radiation protection system was designed for the collider SR and IR to bring the peak power density in the superconducting coils below the quench limit and reduce the dynamic heat deposition in the cold mass by a factor of 100. The system consists of tight tungsten masks in the magnet interconnect regions and elliptical tungsten liners in the magnet aperture optimized for each magnet. These also reduce the background particle fluxes in the collider detector.

# Contents

## Introduction

The discovery of the Higgs boson boosted interest in a low-energy medium-luminosity Muon Collider (MC) as a Higgs Factory (HF). A preliminary design of the 125 GeV c.o.m. HF muon Storage Ring (SR) lattice, the Interaction Region (IR) layout and superconducting (SC) magnets, along with the first results of heat deposition simulations in SC magnets, is described in Refs. [1–3]. It was shown that the large normalized beam emittance and $\beta_{max}$, and the necessity to protect the magnets and detector from showers generated by muon decay products, require very large-aperture SC magnets in both the Interaction Region (IR) and the rest of the ring. A preliminary design of the HF storage ring is based on $Nb_3Sn$ SC magnets with the coil aperture ranging from 50 cm in the interaction region to 16 cm in the arc [2]. The coil cross-sections were chosen based on operating margin, field quality and quench protection considerations to provide an adequate space for the beam pipe, helium channel and inner absorber (liner).

   At the 62.5 GeV muon energy and $2\times10^{12}$ muons per bunch intensity, the electrons from muon decays deposit more than 300 kW in the superconducting magnets of the HF IR and SR [3]. This heat deposition corresponds to an unprecedented average dynamic heat load of 1 kW/m around the 300-m long ring, a multi-MW room temperature equivalent if the heat is deposited at liquid-helium temperature. This paper presents the results of a thorough optimization of the protection system to substantially reduce radiation loads on the HF magnets as well as particle backgrounds in the collider detector.

## 1. Magnets

The main goals of magnet development for MC SR and IR include providing realistic field maps for lattice and IR design analysis and optimization, as well as for studies of beam dynamics and magnet protection against radiation [3, 4].

### 1.1 Design constraints

The rather high magnetic fields and operating margins required for MC SR magnets call for advanced superconductor and accelerator-magnet technologies beyond traditional NbTi magnets. The NbTi superconductor used in all present accelerators has a critical temperature $T_{c0}$ of about 9.8 K and an upper critical magnetic field $B_{c2}$ of about 14.5 T. These parameters limit the operating magnetic fields in accelerator magnets based on this superconductor to 6–7 T at 4.5 K or 8–9 T at 1.9 K. A practical alternative to NbTi is $Nb_3Sn$ superconductor which has a critical temperature $T_{c0}$ of about 18 K and an upper critical magnetic field $B_{c2}$ of about 28 T. Thanks to its superior superconducting properties, $Nb_3Sn$ allows reaching operating fields up to 15–16 T at 4.5 K.

Progress in raising the performance parameters of commercial $Nb_3Sn$ superconducting composite wires in the late 1990s – early 2000s [5] and impressive achievements of accelerator magnet technologies based on this superconductor during the past two decades [6] make it possible to consider $Nb_3Sn$ accelerator magnets for MC storage rings. Due to the significant challenges and uncertainties in operating conditions of superconducting magnets in the MC SR, it is reasonable to choose a nominal operating field at the level of 10 T to provide large (up to 50%) operating margin for MC SR magnets. Conceptual designs and performance parameters of MC SR and IR magnets based on $Nb_3Sn$ superconductor are described below. The coil cross-sections were optimized to achieve the necessary field level and quality in the area occupied by beams using the ROXIE code [7].

The next subsection discusses the conceptual designs and parameters of the HF IR and SR magnets. The lattice parameters are those of Ref. [4]. The large beam size in the IR quadrupoles, as well as the requirements for magnet and detector protection from muon decay showers, leads to very large magnet apertures, which in turn imposes challenging engineering constraints on the magnet design and creates beam dynamics issues with magnet field quality and fringe fields.

### 1.2 Requirements

Table 1 summarizes the IR magnet parameters. The orbit sagitta in the IR dipoles is quite large, 8.1 cm. However, it does not affect the IR dipole bore diameter, which is determined by the large vertical beam size.

Tables 2 and 3 present the maximum values of the main parameters for the dipoles (B) and quadrupoles (Q) used in the chromaticity correction section (CCS), the matching section (MS) and the arc (ARC). The most challenging magnets are the CCS dipoles ($B_{CS}$), some MS dipoles ($B_{MS}I$) and the arc dipoles ($B_{ARC}$) which need high nominal operating field up to 10 T. This field level requires using the $Nb_3Sn$ technology.

The magnet aperture outside the IR reduces from 231 mm in the CCS quadrupoles to 92 mm in the arc dipoles. Note that the aperture size in the arc is defined by the arc dipoles due to a relatively large beam sagitta. To standardize the magnet designs in the various sections it was decided to use for this study two different aperture sizes: large (~231 mm) in the CCS and adjacent part of the MS, and small (~130 mm) in the ARC and adjacent part of the MS.

Table 1. IR magnet specifications.

| Parameter | Q1 | Q2 | Q3 | Q4 | B1 |
|---|---|---|---|---|---|
| 10$\sigma_{max}$ (mm) | 234 | 411 | 339 | 415 | 405 |
| $G_{nom}$ (T/m) | 74 | -36 | 44 | -25 | 0 |
| $B_{nom}$ (T) | 0 | 2 | 0 | 2 | 8 |
| $L_{mag}$ (m) | 1.00 | 1.40 | 2.05 | 1.70 | 4.10 |
| Coil aperture (mm) | 267 | 444 | 372 | 448 | 438 |
| Quantity | 1 | 2 | 1 | 1 | 2 |

**Table 2.** SR dipole magnet parameters

| Parameter | $B_{CS}$ | $B_{MS}I$ | $B_{MS}II$ | $B_{ARC}$ |
|---|---|---|---|---|
| 10$\sigma_{max}$ (mm) | 225 | 222 | 127 | 92 |
| $B_{nom}$ (T) | 10 | 10 | 6.4 | 10 |
| $L_{mag}$ (m) | 1.8 | 2.4 | 3.6 | 3.0 |
| Coil aperture (mm) | 258 | 255 | 160 | 120 |
| Quantity | 13 | 2 | 3 | 8 |

**Table 3.** SR quadrupole magnet parameters.

| Parameter | $Q_{CS}$ | $Q_{MS}$ | $Q_{ARC}$ |
|---|---|---|---|
| 10$\sigma_{max}$ (mm) | 231 | 130 | 46 |
| $B_{nom}$ coil (T) | 5.3 | 5.5 | 3.3 |
| Length (m) | 1.0 | 1.0 | 1.0 |
| Coil aperture (mm) | 264 | 163 | 79 |
| Quantity | 5 | 5 | 4.5 |

The aperture of magnet coils in this analysis was defined as the maximum beam size (10$\sigma_{max}$, i.e. ± 5$\sigma_{max}$ from the beam axis) plus 30 mm for the beam pipe (BP) and absorber (ABS) plus 3 mm for the BP insulation and helium channel. In the IR magnets, the coil aperture was increased by ~50 mm with respect to this definition and limited by two sizes, 320 mm in Q1 and 500 mm in Q2–Q4 and B1–B2, to reserve extra space for IR magnet protection against radiation. The coil aperture of the CCS and adjacent MS magnets was increased to 270 mm, and in the ARC magnets and adjacent MS magnets to 160 mm. This allows using thicker inner absorbers in the arc magnets if necessary.

### 1.3 Storage ring

The coil cross-sections were optimized to achieve the required nominal operating field or gradient with margin and good field quality corresponding to 8$\sigma$ of the full beam size. The conceptual designs of the HF SR magnets are based on a 42-strand Rutherford cable, 21.6 mm wide and 1.85 mm thick, made of a 1 mm strand with Cu/nonCu ratio of 1.0. The cable is insulated with a 0.2 mm thick insulation. All magnets are based on 2-layer, shell-type coils with the iron yoke used mainly to reduce the fringe fields.

The main magnet parameters at 4.5 K are reported in Table 4. The relatively low fields in CCS, MS and ARC quadrupoles allow using the traditional NbTi technology. Therefore, the parameters of these magnets were calculated for both conductor options, NbTi and Nb$_3$Sn.

**Table 4.** Nominal magnet parameters at $T_{op}$ = 4.5 K.

| Parameter | Dipole | | Quadrupole | |
|---|---|---|---|---|
| Coil ID (mm) | 270 | 160 | 270 | 160 |
| Max $B_{op}$ (T) | 10.0 | 10.0 | - | - |
| $B_{max}$ (T) | 14.1 | 14.2 | - | - |
| Max $G_{op}$ (T/m) | - | - | 33 | 36 |
| $G_{max}$ (T/m) | - | - | 54.1[*]/96.6[#] | 90.0[*]/160.8[#] |
| $B_{max}$ coil (T) | 15.8 | 15.6 | 8.7[*]/15.3[#] | 8.3[*]/14.7[#] |
| Fraction of SSL | 0.71 | 0.71 | 0.61[*]/0.34[#] | 0.4[*]/0.22[#] |
| $L$ (mH/m) | 54.1 | 20.5 | 24.5 | 8.7 |
| $E_{op}$ (MJ/m) | 4.33 | 1.84 | 0.59 | 0.12 |
| $F_{x\_op}$ (MN/m) | 9.25 | 5.98 | 0.89 | 0.25 |
| $F_{y\_op}$ (MN/m) | -3.96 | -2.66 | -0.93 | -0.28 |

[*] NbTi coil.
[#] Nb$_3$Sn coil.

The optimized D and Q coil cross-sections for CCS, MS and ARC sections are shown in Fig. 1. The Nb$_3$Sn dipole magnets operate at 71% and the quadrupoles at 22–34% of their short sample limit (SSL) at 4.5 K. Using the same cable with NbTi strands in both quadrupoles reduces $G_{max}$ to 54 and 90 T/m. However, the magnets still have a sufficient operating margin as they operate at 61% and 40% of the SSL for the 270 mm and 160 mm quads respectively. The final choice of superconductor for the SR quadrupoles will depend on the results of radiation studies for these magnets. Geometric field harmonics at the corresponding reference radius are reported in Table 5. With optimized coil cross-sections the relative field errors in the area occupied by muon beams is ~10$^{-4}$ (dark blue area in Fig. 1).

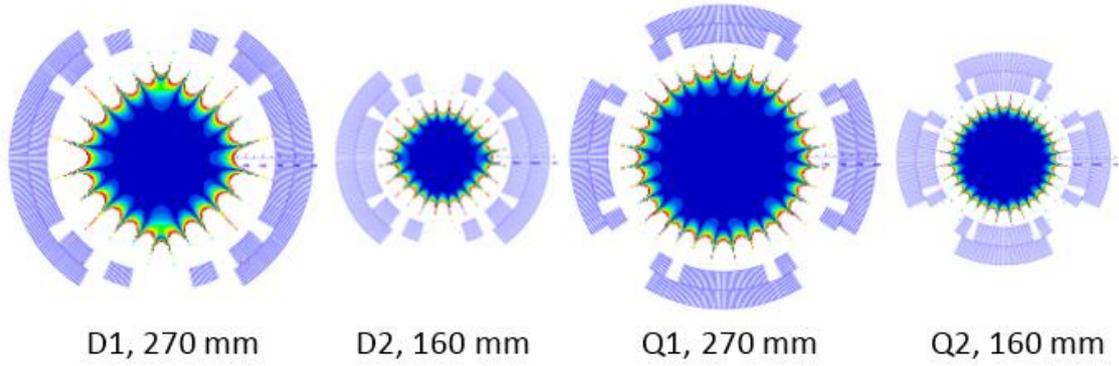

**Figure 1.** Coil cross-sections of HF SR dipoles and quadrupoles.

**Table 5.** Geometric harmonics at $R_{ref}$ ($10^{-4}$).

|   | ID, mm | $R_{ref}$, mm | $b_3$ | $b_5$ | $b_6$ | $b_7$ | $b_9$ | $b_{10}$ |
|---|---|---|---|---|---|---|---|---|
| D | 270 | 90 | -0.2 | -0.1 | - | 0.2 | 1.2 | - |
|   | 160 | 53 | 0.0 | -0.1 | - | 0.5 | 1.1 | - |
| Q | 270 | 90 | - | - | 0.2 | - | - | 0.0 |
|   | 160 | 53 | - | - | 0.1 | - | - | 0.0 |

**1.4 Interaction region**

Conceptual designs of the HF IR magnets are based on a 1 mm Nb$_3$Sn strand with J$_c$(12T,4.2K) of 2.7 kA/mm$^2$ and Cu/nonCu ratio of 1.15. The Q1–Q4 and B1 coils use a 42-strand Rutherford cable, 21.6 mm wide and 1.85 mm thick. The cable in the dipole coil Bq (in Q2 and Q4) has 22 strands and is 11.3 mm wide and 1.77 mm thick. Both cables are insulated with a 0.2 mm thick insulation. The volumes of all coils were chosen based on operating margin and quench protection considerations. The coil aperture of the IR magnets was increased by 50 mm, to 320 mm in Q1 and to 500 mm in Q2–Q4 and B1, to provide adequate space for the beam pipe, helium channel and inner absorber (liner). The coil cross-sections were optimized to achieve the required nominal gradient or dipole field with sufficient operating margin and good field quality in the beam area.

The IR magnets use 6-layer, shell-type coils. The dipole coil Bq in Q2 and Q4 has only one layer. Q1 and Q2(1) do not have an iron yoke since they operate inside the detector field. Q2(2)–Q4 and B1 use the iron yoke mainly to reduce fringe fields. The optimized cross-sections of the Q1–Q4 and B1 coils are shown in Fig. 2. The main magnet parameters at 4.5 K are reported in Table 6. The parameters for Q2 and Q4 (with dipole coil Bq) include the combined dipole and quadrupole fields. The Bq parameters include the field from the main quadrupole coil Q2 at $G_{nom}$ = 35 T/m. In Q4, the Bq parameters are better due to the lower $G_{nom}$ = 25 T/m.

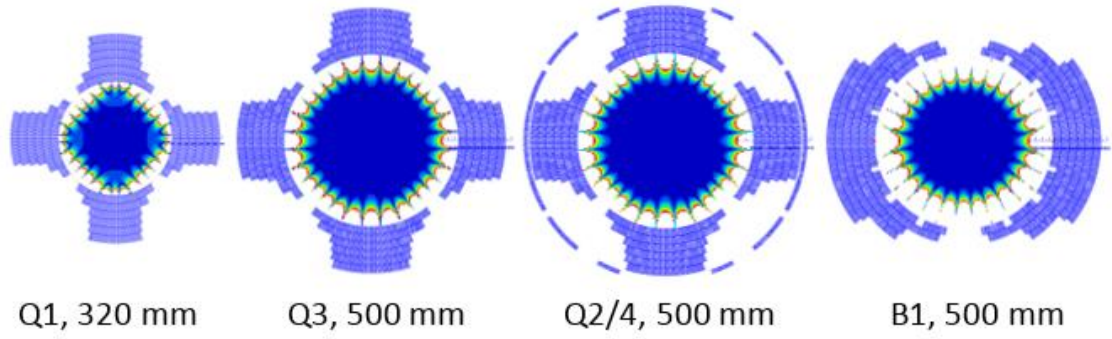

Q1, 320 mm    Q3, 500 mm    Q2/4, 500 mm    B1, 500 mm

**Figure 2.** Coil cross-sections of IR magnets: Q1, Q3, Q2 and Q4 with dipole coil Bq and B1 dipole.

**Table 6.** IR magnet parameters at $T_{op}$ = 4.5 K

| Parameter | Q1 | Q2* | Q3 | Q4* | Bq** | B1 |
|---|---|---|---|---|---|---|
| Aperture (mm) | 320 | 500 | 500 | 500 | 780 | 500 |
| $B_{max}$ (T) | - | - | - | - | 4.18 | 16.3 |
| $G_{max}$ (T/m) | 94.9 | 58.2 | 62.9 | 58.2 | - | - |
| $B_{t\_max}$ coil (T) | 16.4 | 17.2 | 16.9 | 17.2 | 15.0 | 17.7 |
| Operating margin | 0.78 | 0.62 | 0.70 | 0.62 | 0.48 | 0.50 |
| $L$ (mH/m) | 177 | 454 | 454 | 454 | 65 | 1188 |
| $E_{op}$ (MJ/m) | 10.4 | 11.3 | 16.8 | 5.5 | 1.6 | 13.8 |
| $F_x(I_{op})$ (MN/m) | 5.8 | 7.3 | 6.5 | 4.1 | 0.7 | 9.7 |
| $F_y(I_{op})$ (MN/m) | -12.3 | -12.6 | -14.2 | -6.8 | -4.7 | -8.9 |

* calculated for nominal Bq = 2 T.
** calculated for $G_{nom}$ = 35 T/m (Q2).

    The operating margin is defined as the ratio of magnet nominal to maximum field or field gradient. A 6-layer coil design provides sufficient operating margin in the IR magnets at relatively low current density in the coil. All the magnets operate at 50–80% of the short sample limit at 4.5 K. The quadrupole and dipole coils also operate at a high level of Lorentz forces, which requires stress management in the azimuthal direction. Conductor grading in the coil will provide additional margin, if needed.

    Geometric field harmonics for Q1–Q4, Bq and B1 at corresponding $R_{ref}$ are shown in Table 7. Coil cross-section optimization provided relative field errors in the area occupied by muon beams at the level of $10^{-4}$ (dark blue area in Fig. 2). Analysis of the Dynamic Aperture (DA) with the MADX PTC code shows that field errors in the straight sections of the IR magnets reduce the DA by a factor of 2 so that it coincides with the good field region shown in Fig. 2.

Table 7. Geometric Harmonics at $R_{ref}$ ($10^{-4}$).

| Parameter | Q1 | Q2–Q4 | Bq | B1 |
|---|---|---|---|---|
| $R_{ref}$ (mm) | 135 | 225 | 225 | 225 |
| $b_3$ | - | - | 0.08 | -0.07 |
| $b_5$ | - | - | 0.05 | -0.06 |
| $b_6$ | -0.56 | -0.18 | - | - |
| $b_7$ | - | - | 0.34 | 0.12 |
| $b_9$ | - | - | 0.57 | 0.02 |
| $b_{10}$ | -0.47 | -0.57 | - | - |
| $b_{11}$ | - | - | 2.55 | 0.91 |
| $b_{13}$ | - | - | -2.89 | 1.58 |
| $b_{14}$ | 3.45 | -0.94 | - | - |

**1.5 Implementation in MARS15 model**

The MARS15 Monte Carlo code [8] is used to address the key issues related to magnet and detector protection from radiation in the HF MC. A detailed 3D model of the entire collider ring including the IR, chromaticity correction (CCS) and matching (MS) sections, arc, machine–detector interface (MDI), as well as the SR tunnel and detector hall has been built [3] with described magnet geometry, materials and magnetic fields. Figures 3 and 4 show the model of the 300-m circumference HF with the SiD-like detector at the interaction point (IP). The silicon vertex detector and tracker are based on the design proposed for the CMS detector upgrade. The detector geometry in GDML format was promptly imported into the MARS15 model.

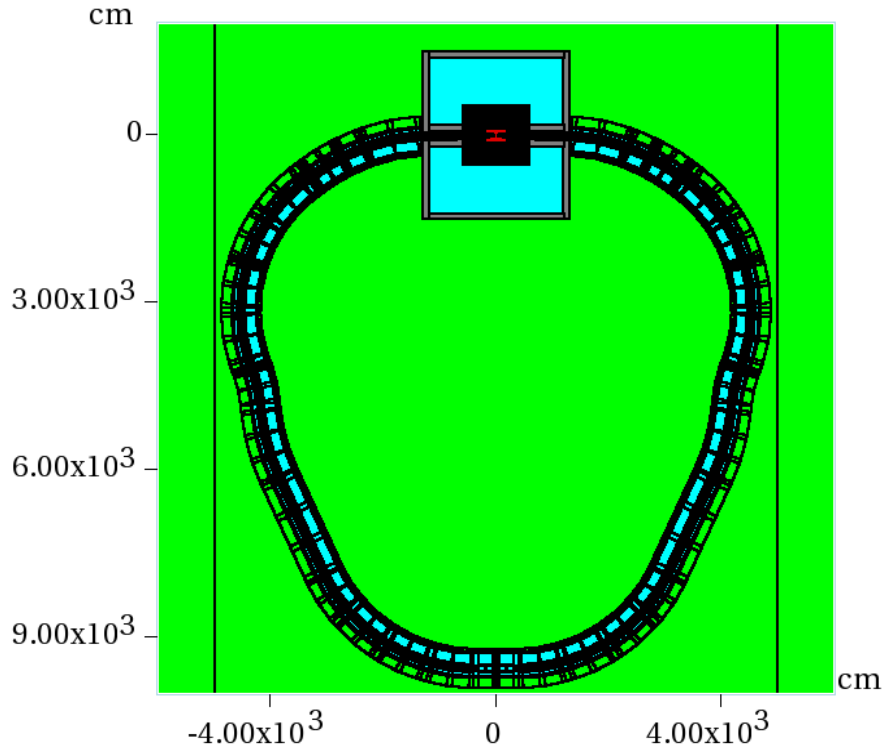

**Figure 3.** MARS15 model of the Higgs Factory Muon Collider.

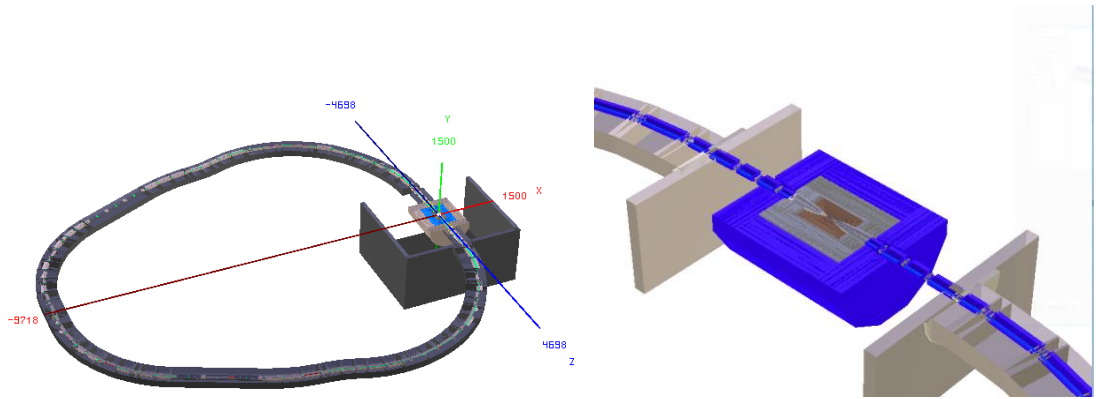

**Figure 4.** 3D MARS15 models of the HF MC (left) and MDI with SiD-like detector (right).

Cross-sectional views of the IR combined-function quadrupoles Q2 and Q4 as designed and implemented in the MARS15 model are presented in Fig. 5, while Fig. 6 shows those for the 50-cm ID 8-Tesla IR dipoles.

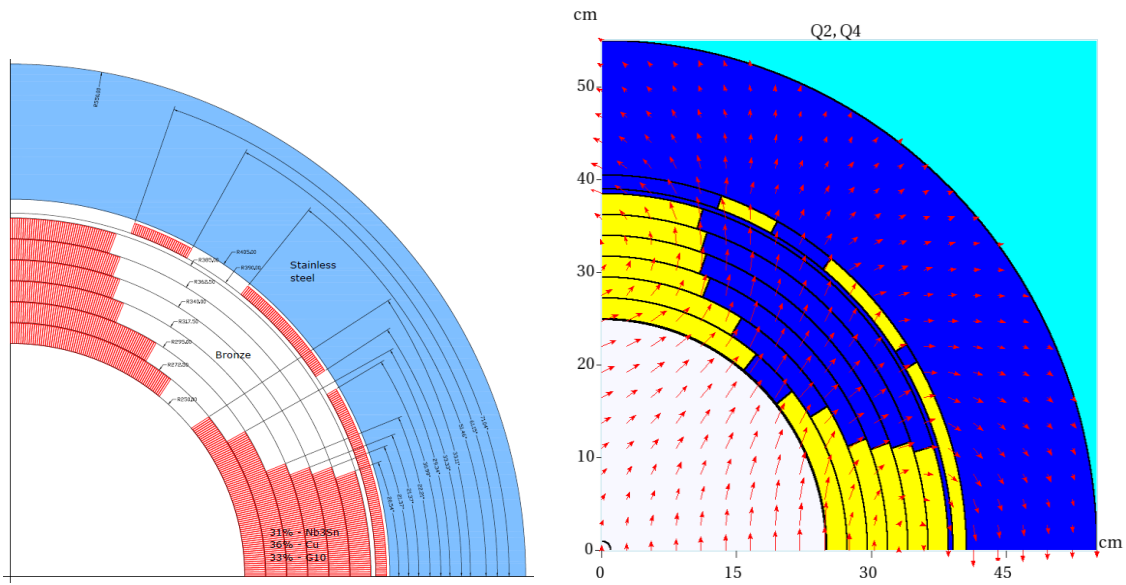

**Figure 5.** 50-cm ID IR quadrupoles Q2 and Q4 as designed (left) and built-in MARS15 model (right).

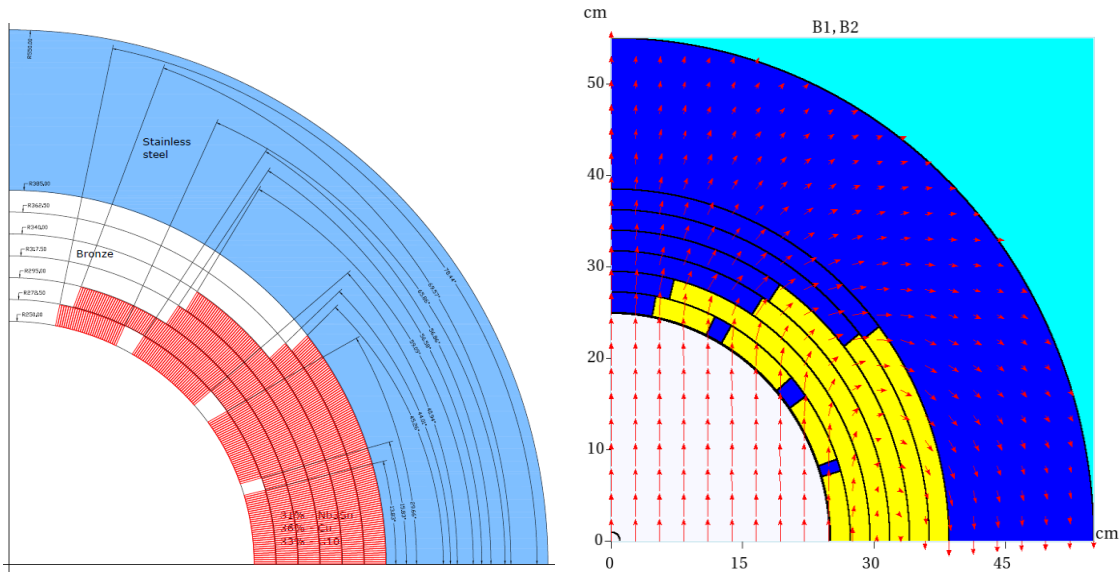

**Figure 6.** 50-cm ID 8-Tesla IR dipoles B1 and B2 as designed (left) and built-in MARS15 model (right).

## 2. Magnet protection system

### 2.1 Concept

The first studies of radiation heat depositions and doses in the HF collider ring have shown [1, 2] that to provide the required operational stability and adequate lifetime of the HF SC magnets the values of radiation load should be reduced by at least a factor of 100. It was shown in early studies [9, 10] that the practical way to protect SC magnets of a muon collider ring against electromagnetic showers induced by electrons from muon decays is to
- Limit the magnet lengths to 2 to 4 m;
- Install tight tungsten masks in the magnet interconnect regions;
- Place thick tungsten liners inside the magnet apertures.

Such a radiation protection system concept allows reaching the following goals:
1. Provide reduction of the peak power density in the $Nb_3Sn$ cable to ~1.5 mW/g which is below the $Nb_3Sn$ superconductor quench limit with an appropriate safety margin;
2. Keep the lifetime peak dose in the innermost layers of insulation below 20-40 MGy;
3. Reduce the average dynamic heat load in the cold mass to the level of ~10 W/m, acceptable for a cryogenic system;
4. Suppress the long-range component of the detector background.

As a result of massive MARS15 simulations, such a magnet protection system (MPS) has been designed for the HF MC to fulfil these constraints and to provide at least a $8\sigma_{x,y}$ full beam envelope ($\pm 4\sigma_{x,y}$ from the beam axis) for muons for up to 2000 turns.

### 2.2 Reducing heat load on cold mass

The protection system parameters have been individually optimized for each magnet and interconnect region in the 300-m circumference HF collider ring and IR. Figure 7 presents a fragment of the radiation protection system built in MARS15 for the IR. As an example, the thickest tungsten liner for one of the hottest dipole magnets in the CCS at 24.2 m from the IP is shown in Fig. 8. The liner is 4.1 cm thick horizontally and 2 cm thick vertically and is cooled to 60–80 K. It is followed radially by the support structure, stainless steel beam pipe, Kapton insulation, liquid helium channel and 4.5 K Nb$_3$Sn coil.

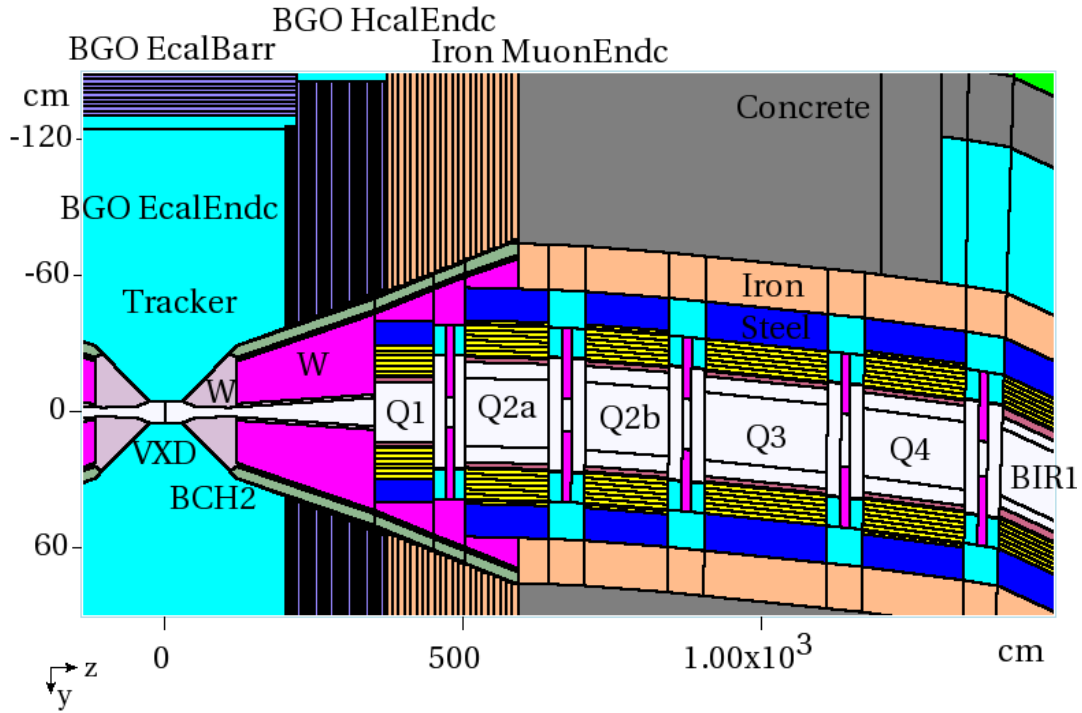

**Figure 7.** MDI MARS15 model with tungsten nozzles on each side of IP, tungsten masks in interconnect regions and tungsten liners inside each magnet.

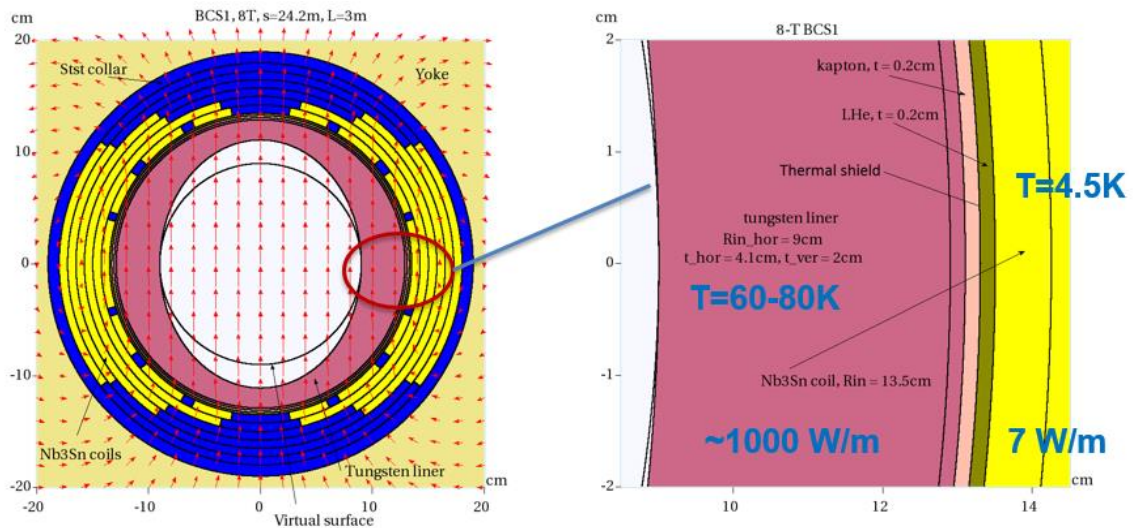

**Figure 8.** Elliptical tungsten liner inside BCS1 8 T dipole.

15-cm long tungsten masks are installed in every interconnect region around the machine (see, e.g., Fig. 7). Their apertures are at ± 4$\sigma_{x,y}$ from the beam axis or farther. Figure 9 (left) shows the geometrically tightest mask at the IP end of the hottest BCS1 dipole at 24.2 m from the IP. Calculated isocontours of the power density at its longitudinal maximum in this BCS1 dipole are shown in Fig. 9 (right). The shielding effect of the tungsten mask and liner in the magnet aperture is clearly seen. The peak power density in the BCS1 dipole SC coil is reduced to less than 1 mW/g from about 150 mW/g at the inner radius of the tungsten liner. The reduction to the target value of 1.5 mW/g or less is achieved in the IR magnets (Fig. 10) as well as in all the SC coils around the ring.

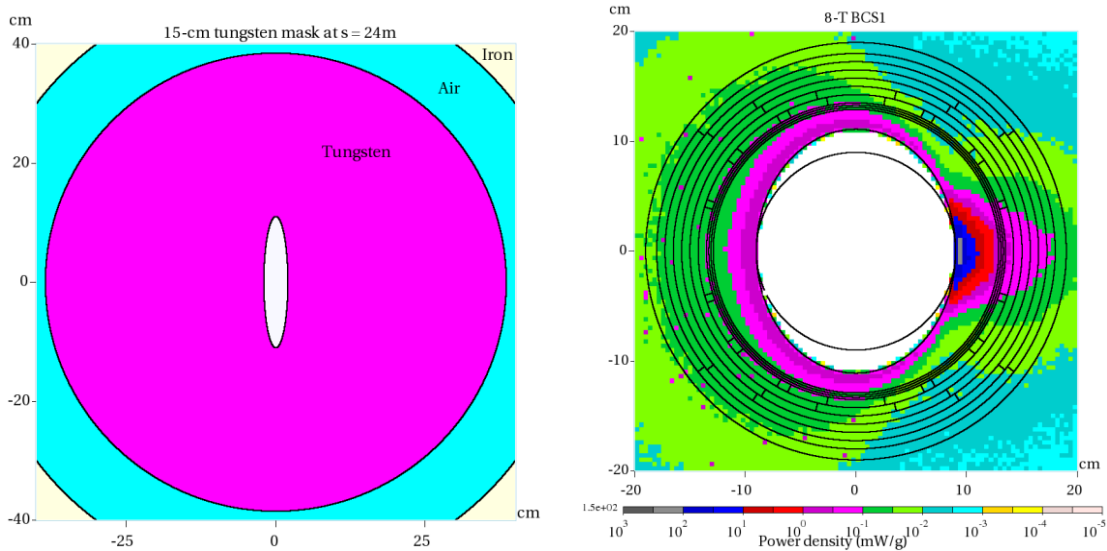

**Figure 9.** Tungsten mask at the IP end of the BCS1 8-T dipole (left) and power density isocontours in this dipole (right). The ring center is to the right in these figures.

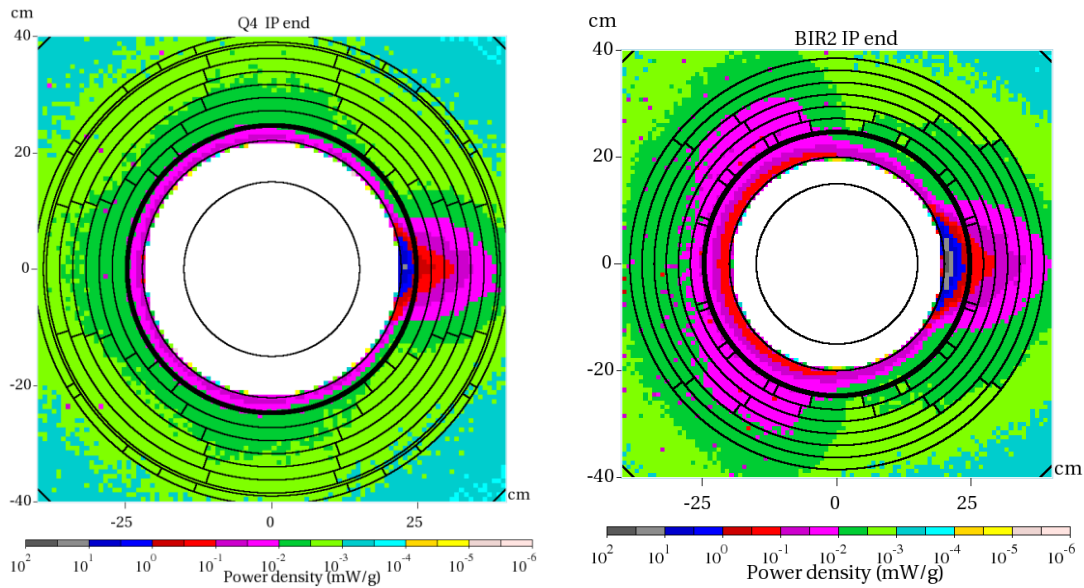

**Figure 10.** Power density isocontours at the IP ends of the Q4 quadrupole (left) and BIR2 dipole (right). The ring center is to the right in these figures.

Dynamic heat loads on the SC magnet cold mass define the capacity of the collider cryoplant and its operating costs. An acceptable level of dynamic heat load is 10 W/m or less at the 4.5 K liquid helium temperature. This means that the magnet protection system needs to reduce the average load to the cold mass by factor of 100 from the original 1 kW/m. The system described above was designed under this constraint, and results are shown in Fig. 11. The average load to the cold mass is below the desired 10 W/m value. Although in some magnets the heat load level is up to 15 W/m which is higher than average, it is still tolerable for a cryogenic system at 4.5 K temperature. The remaining average power dissipation of ~990 W/m is intercepted by the tungsten masks and liners cooled to 60–80 K.

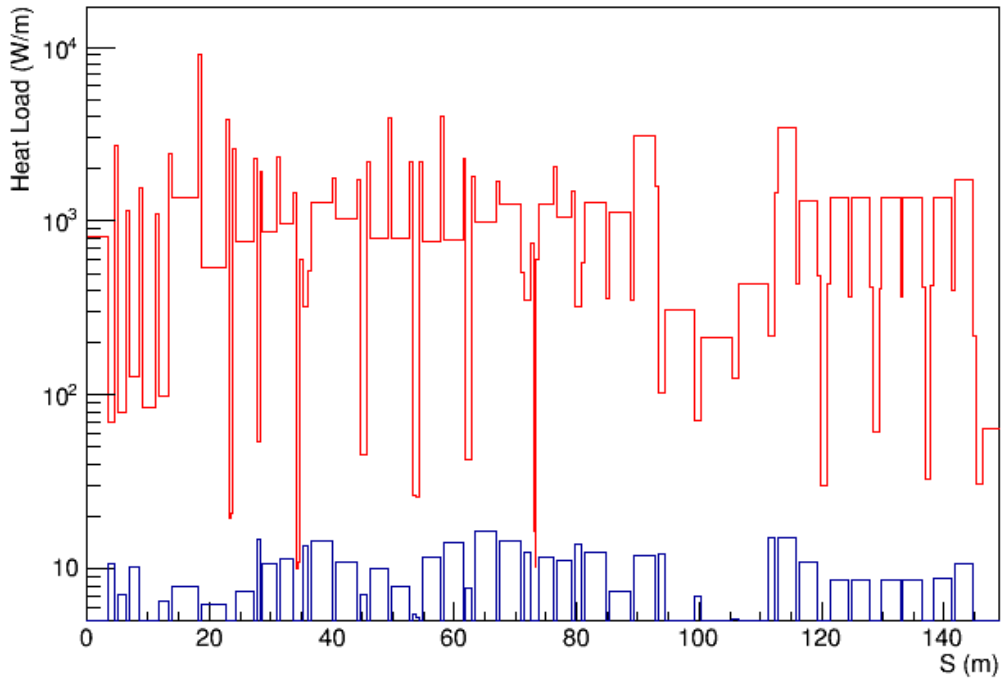

**Figure 11.** Dynamic heat load in SC magnets along half of the HF collider for tungsten masks and liners at 60–80 K (red) and cold mass at 4.5 K (blue).

The designed tungsten masks and liners are positioned at $\pm 4\sigma_{x,y}$ from the beam axis or farther. The analysis of related resistive wall impedance and beam stability has shown [11] that one can expect some small (few percent) growth of an initial perturbation after 1000 turns, so there is a safety factor of about one hundred for transverse-plane instabilities. For the longitudinal plane, the magnet protection system can result in up to ~30% energy broadening. This effect could probably be mitigated by means of second-harmonic RF. In any case, thin conducting tapers will be included in the system for a smooth transition between masks and liners.

## 2.3 Machine–detector interface

The above MPS design also helps reduce long-range background particle load on the HF Muon Collider detector [12]. It includes an additional crucial element: a nozzle inside the detector to intercept products of shower development in the IP vicinity. Figure 12 shows that — thanks to the MPS described in this paper — only energetic photons and Bethe–Heitler muons come to the nearest IP Q1 quadrupole at $z = 4.4$ m. The nozzle design and its effect in background particle flux reduction in the detector is described in Ref. [12].

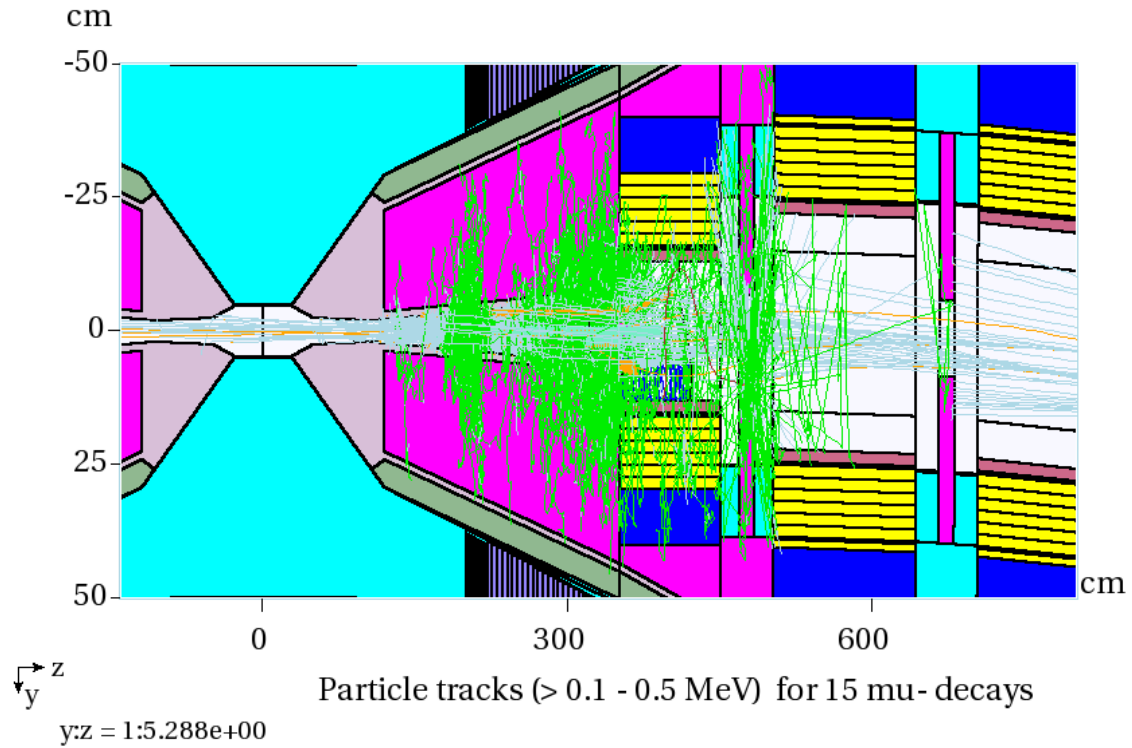

**Figure 12.** Visualization of the machine MPS and nozzle particle background reduction at IP.

## 3. Conclusions

Superconducting magnet designs have been proposed for the Higgs Factory Muon Collider. These provide high operating gradient and magnetic field in a large aperture to accommodate the large size of muon beams as well as the cooling system to intercept the large heat deposition from the showers induced by decay electrons. The magnet geometry, materials and magnetic fields are implemented in a detailed 3D MARS15 model of the entire HF collider ring including IR, chromaticity correction and matching sections, arc, and machine–detector interface. A sophisticated radiation protection system based on tight tungsten masks in the magnet interconnect regions and optimized elliptical tungsten liners in the magnet apertures was designed for the HF collider. This system allows reduction of the peak power density in the superconducting coils to below their quench limit and the dynamic heat deposition in the magnet cold masses to a level tolerable by the cryogenic system. The results obtained confirm the possibility of radiation protection of $Nb_3Sn$ superconducting magnets in a Higgs Factory muon collider.

## Acknowledgments

This document was prepared using the resources of the Fermi National Accelerator Laboratory (Fermilab), a U.S. Department of Energy, Office of Science, HEP User Facility. Fermilab is managed by Fermi Research Alliance, LLC (FRA), under Contract No. DE-AC02-07CH11359. The authors are thankful to Y. Alexahin, A. Burov and D. Kaplan for useful discussions.